\documentclass[twocolumn,aps,prl,preprintnumbers]{revtex4}
\usepackage{bbm}
\usepackage[latin9]{inputenc}
\usepackage{amsmath}
\usepackage{amssymb}
\usepackage{empheq}
\usepackage{graphicx}
\usepackage{mathrsfs}
\usepackage{amsfonts}
\usepackage{amsthm}
\usepackage{color}
\usepackage{txfonts}
\usepackage{ulem}
\providecommand{\U}[1]{\protect\rule{.1in}{.1in}}
\makeatletter
\@ifundefined{textcolor}{}
{
\definecolor{BLACK}{gray}{0}
\definecolor{WHITE}{gray}{1}
\definecolor{RED}{rgb}{1,0,0}
\definecolor{GREEN}{rgb}{0,1,0}
\definecolor{BLUE}{rgb}{0,0,1}
\definecolor{CYAN}{cmyk}{1,0,0,0}
\definecolor{MAGENTA}{cmyk}{0,1,0,0}
\definecolor{YELLOW}{cmyk}{0,0,1,0}
}

\makeatother
\begin{document}
\title{Twisted skyrmion at domain boundaries and the method of image}
\author{Huanhuan Yang}
\author{C. Wang}
\email[corresponding author: ]{cwangad@connect.ust.hk}
\author{Xiaofan Wang}
\author{X.S. Wang}
\author{Yunshan Cao}
\author{Peng Yan}
\email[corresponding author: ]{yan@uestc.edu.cn}
\affiliation{School of Microelectronics and Solid-state
Electronics and State Key Laboratory of Electronic Thin Film and
Integrated Devices, University of Electronic Science and Technology
of China, Chengdu 610054, China}
\date{\today}

\begin{abstract}
We predict a novel twisted skyrmion structure at
the boundary of two antiferromagnetically coupled 
magnetic domains with antiparallel magnetization 
directions. Through this intermediate state, 
skyrmions with opposite polarities can be freely 
switched between each other by spin-polarized electric currents.
Based on these findings, we propose the concept of 
double-track skyrmion racetrack memory and logic gates 
where the binary data are represented by skyrmions
with different polarities. The dynamics of skyrmion 
polarity reversal is theoretically studied.
Using the method of image, we derive the analytical 
formula of the repulsive potential when a normal skyrmion approaches
the domain boundary. A harmonic attractive potential well is
obtained for the twisted skyrmion across the boundary.
Micromagnetic simulations compare well with theoretical predictions.
The method of image skyrmion proposed in this work can be used
to deal with a large class of skyrmion-boundary interaction problems.
\end{abstract}

\maketitle

\emph{Introduction.$-$}Magnetic skyrmions are stable swirling noncoplanar
topological defects in the magnetization texture. Since the theoretical
prediction more than two decades ago \cite{Bogdanov1,Bogdanov2} and the
recent experimental observation \cite{Muehlbauer}, intensive investigation
on magnetic skyrmions gives birth to an emerging subfield of condensed matter
physics, the skyrmionics \cite{Roland,Fert,Wanjun,SeKwon}.
Due to their small size, excellent stability and low driving
current, skyrmions are promising information carriers in future ultra-dense
(and -fast) spintronic applications, of which the most attractive proposals
are skyrmion racetrack memories (SRMs) and skyrmion logic gates (SLGs), 
while data representation is still a key obstacle. Unlike their domain-wall 
counterpart \cite{Parkin,Cowburn} where the binary data bits ``1" and ``0"
are encoded by magnetic domains of opposite spin directions
integrated with the magnetoresistive (MR) sensor, it is impossible
to reverse the core polarity of a single skyrmion on the
track, unless all spins of the magnetic domain are switched \cite{Heo,Zeng}.
The topologically protected stability of magnetic skyrmions is thus
double-edged. A compromised approach is to represent ``1" and ``0" by
the presence and absence of a skyrmion (or the other way around),
respectively. This method, however, requires an exactly synchronized
motion of all skyrmions. If there is a pinning or any relative motion
between the skyrmions, errors happen inevitably, and one cannot identify
the errors from normal data. Although there are proposals of two-lane SRM
in the literature claimed to overcome the problem \cite{Kang,Muller1},
a reliable data representation is still not achieved since those proposals
rely on the presence and absence of skyrmions, instead of skyrmions 
carrying opposite polarities.
\par

\begin{figure}[ht!]
	\centering
	\includegraphics[width=0.45\textwidth]{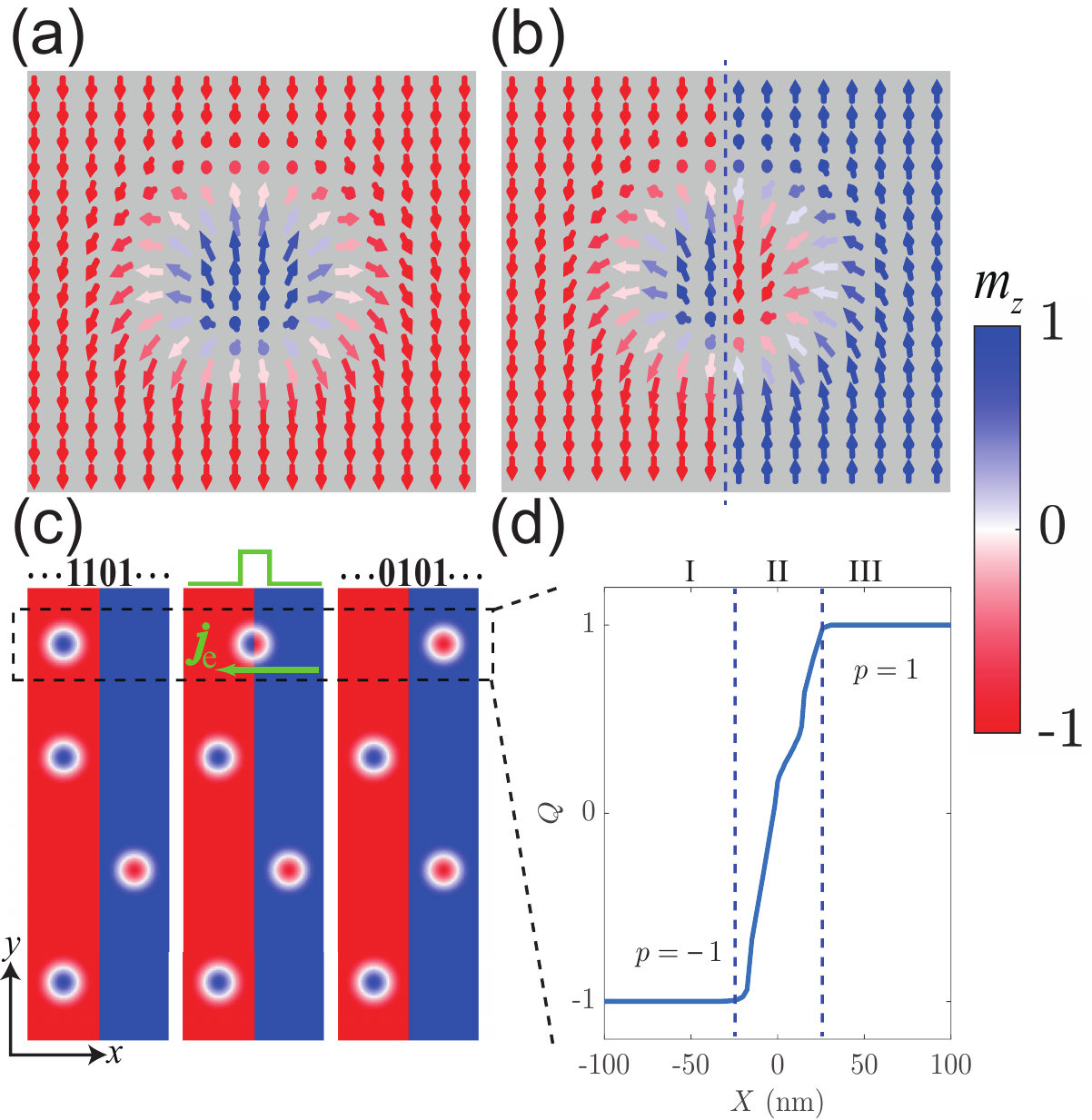}
	\caption{(a)(b) Illustration of a
    normal skyrmion in a single magnetic domain (a) and a twisted
    skyrmion at the boundary (dashed line) of two antiparallel
    magnetic domains (b).
    (c) Schematics (top view) of four bits consisting of four
    skyrmions in a double-track SRM. A local electric current
    (green arrow) is injected along $x-$axis (electrons move from
    left to right) to drive the first skyrmion across the boundary,
    and to flip its core polarity at the same time, so that the
    binary data ``1101" is rewritten to be ``0101".
    (d) Position dependence of the skyrmion charge $Q$. Its
    polarity $p$ changes the sign after the skyrmion passes
    through the boundary.}
	\label{fig1}
\end{figure}

Tantalizing physics often emerges at boundaries, such as
the recently discovered quantum spin Hall insulator
states \cite{TI} and chiral majorana fermion modes \cite{He},
among others. As to magnetic skyrmions, the edge effect
in most cases is viewed as harmful because the intrinsic
skyrmion Hall effect tends to result in skyrmion
accumulations there. Even worse, skyrmions can be annihilated
by the edge. How skyrmions interact with the boundary is not solved
in a satisfactory manner in the community. In this Letter, we
predict a novel stable twisted skyrmion state at the boundary of two
antiferromagnetically (AFM) coupled magnetic domains aligned
in an antiparallel way (shown in Fig.~\ref{fig1}). Through this
intermediate state, skyrmions in the two domains can be freely
switched between each other by, for example, applying an electric current.
This enables us to design a double-track SRM, as well
as SLGs, in which skyrmions with opposite polarities are utilized 
to encode ``1'' and ``0''. Our proposal removes the mentioned barrier 
for achieving the SRM and SLG.
Furthermore, using the method of image, we derive the analytical formula of the
potential energy when a normal skyrmion approaches the domain boundary.
Our analytical results show that the potential is repulsive and
inversely proportional to the square of the distance between the
skyrmion and the boundary in the large distance limit, while it
becomes linear when the skyrmion is very close to the boundary.
After the skyrmion steps into the domain boundary, a twisted
skyrmion forms and is confined by a harmonic attractive potential
well. Micromagnetic simulations compare very well with theoretical
predictions. The method of image skyrmion we developed can be used
to solve a large class of skyrmion-boundary interaction problems.

\emph{Model.}$-$We start from the following classical spin
Hamiltonian in two spatial dimensions ($x-y$ plane),
\begin{equation}
\begin{gathered}
\mathcal{H}=-\sum_{\langle ij \rangle\not\in\mathrm{B}}
J_{\text{FM}}(\mathbf{m}_i \cdot \mathbf{m}_j)
+\sum_{\langle ij \rangle\in\mathrm{B}}J_{\text{AFM}}
(\mathbf{m}_i\cdot\mathbf{m}_j)\\
-\sum_{\langle ij\rangle}
\mathbf{D}_{ij}\cdot(\mathbf{m}_i\times
\mathbf{m}_j)
-\sum_{i} K(\mathbf{m}_i\cdot \hat{z})^2
+\mathcal{H}_{\text{DDI}}.
\end{gathered}\label{Hamiltonian_1}
\end{equation}
Here $\mathbf{m}_i$ is the unit vector of spin
at the site $i=(i_x a,i_y a)$ ($i_{x(y)}$ is an integer
and $a$ is the lattice constant) with saturation magnetization
$M$. $\langle ij\rangle\not\in\mathrm{B}$
($\langle ij\rangle\in\mathrm{B}$) sums over all
nearest-neighbour sites away from (at) the boundary.
$J_{\text{FM}}>0$ is the ferromagnetic exchange constant
in magnetic domains, and $J_{\text{AFM}}>0$ is the AFM
exchange coupling at the boundary of two domains with
antiparallel-aligned magnetizations [see Fig.~\ref{fig1}(b)]. 
The third term in the right hand of Eq. (\ref{Hamiltonian_1}) 
represents the interfacial Dzyaloshinskii-Moriya interaction (DMI) with a
homogeneous DMI vector $\mathbf{D}_{ij}
=D\hat{r}_{ij}\times\hat{z}$, where $\hat{r}_{ij}$
is the unit vector connecting sites $i$ and $j$. Parameter $K>0$
is the easy-axis uniaxial anisotropy constant
along $z-$direction. $\mathcal{H}_{\text{DDI}}$ is the dipole-dipole
interaction energy. A N\'{e}el-type skyrmion is favored in
this model Hamiltonian.

\emph{Creation of the twisted skyrmion.$-$}By
applying electric current, we can push a normal N\'{e}el
skyrmion [shown Fig.~\ref{fig1}(a)] into the domain
boundary where the skyrmion structure is strongly
twisted due to the large interfacial AFM interaction
[shown Fig.~\ref{fig1}(b)]. To test this idea,
we have performed micromagnetic simulations with
the MuMax3 package \cite{mumax}, using materials
parameters of cobalt$|$platinum (Co$|$Pt) \cite{Sampaio}
with the DMI $D=1.4~\hbox{meV}$, the ferromagnetic
exchange constant $J_{\text{FM}}=94~\hbox{meV}$, the
uniaxial anisotropy $K=0.07~\hbox{meV}$, the saturation
magnetization $M=5.8\times 10^5$ A m$^{-1}$, the Gilbert
damping constant $\alpha=0.3$, and the boundary AFM
energy $J_{\text{AFM}}=0.05-3$ $J_{\text{FM}}$. We
consider a sample of size $600\times 2000\times 1~\hbox{nm}^3$
with the lattice constant $a=1~\hbox{nm}$.

One example of twisted skyrmion creation and its dynamics 
is depicted in Fig.~\ref{fig1}(c).
The boundary separating the two domains is at $x=0$. We
choose an interfacial AFM constant $J_{\text{AFM}}=J_{\text{FM}}$, and
consider a single N\'{e}el skyrmion with
polarity $p=-1$ initially in the left magnetic domain (the skyrmion
polarity is defined as follows: $p=\mp1$ corresponds to the core
spin pointing up and down, respectively).
Then, we inject a very large in-plane electric current
$\mathbf{j}_{e}=-j_{e}\hat{x}$ with $j_{e}=10^{13}$ A m$^{-2}$,
to drive the skyrmion to propagate toward the boundary.
The skyrmion crosses the boundary and transforms into a
twisted skyrmion, as shown in the middle panel of Fig.~\ref{fig1} (c).
We find that the twisted skyrmion stops at the boundary if the 
current is turned off at this moment. This result indicates that 
the twisted skyrmion is also a stable magnetic soliton.
As the electric current is still on, the twisted skyrmion 
will keep moving and leave the boundary. Finally it recovers 
the normal state in the opposite domain, however, with a 
reversed polarity $p=1$, as shown in the right panel of 
Fig.~\ref{fig1}(c). Spin transfer torques under a different 
configuration, i.e., current perpendicular to plane, as well 
as the spin-orbit/spin Hall torques originating from the 
relativistic spin orbital coupling can also be adopted to 
drive the skyrmion propagation and its polarity switching \cite{supp}.
Consequently, we can use $p=\mp 1$ to encode the binary
bits ``1'' and ``0'' in a double-track SRM, as shown in
Fig.~\ref{fig1} (c). This encoding approach is quite robust
against external frustrations. For instance, if one skyrmion
is pinned, the double-track SRM with integrated MR sensors
is able to distinguish the error from normal data \cite{supp}.
Skyrmion logic based on the novel data representation concept 
is also proposed and validated \cite{supp}.
\par

Figure~\ref{fig1}(d) shows the topological charge
$Q=\int q d\mathbf{r}$ as a function
of the position of the skyrmion (twisted skyrmion) 
center plotted in Fig.~\ref{fig1}(c).
Here $q(\mathbf{r})=\mathbf{m}\cdot(\partial_x
\mathbf{m}\times\partial_y \mathbf{m})
/(4\pi)$ is the skyrmion charge density expressed
in the continuum limit.
The skyrmion (twisted skyrmion) center is defined as
$\mathbf{X}=(X,Y)\equiv Q^{-1}\int q\mathbf{r} d\mathbf{r}$.
We find that there exist two plateaus, $Q=-1$ and $1$,
corresponding to the state that the skyrmion is in the left and
right domains with $p=-1$ and $1$, respectively.
For the twisted skyrmion state, $Q$ changes continuously 
with $X$.

\begin{figure}[ht!]
    \centering
    \includegraphics[width=0.45\textwidth]{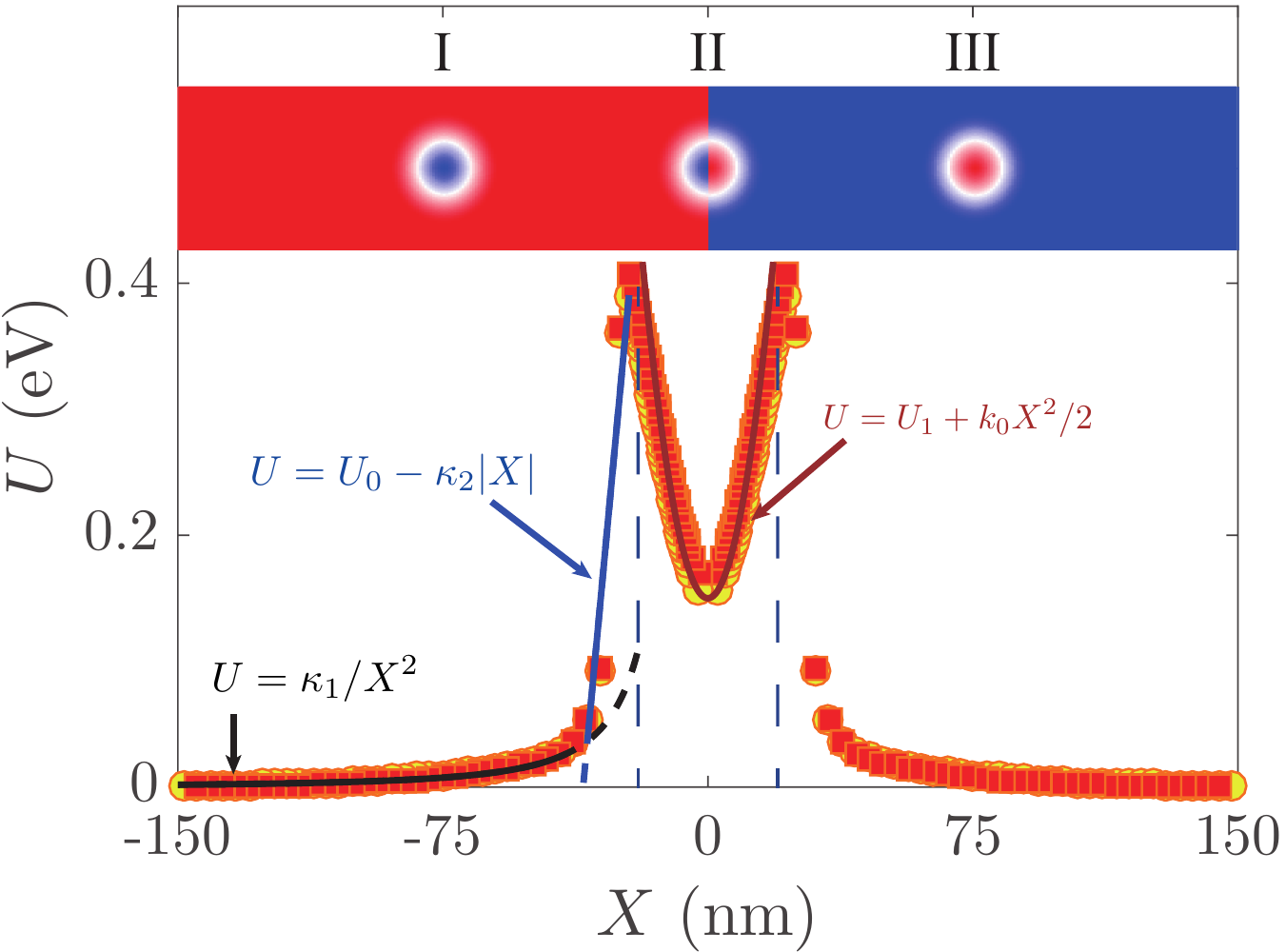}
    \caption{Potential energy $U$ v.s. skyrmion/twisted skyrmion center position
    $X$ for $J_{\text{AFM}}/J_{\text{FM}}=1$ (yellow circles)
    and 3 (red squares).
    The solid black curve is Eq.~\eqref{sky_ima_1} with
    $R= 12~\hbox{nm}$.
    The solid blue line is Eq.~\eqref{sky_ima_2} with
    a smaller $R= 5~\hbox{nm}$ due to the size shrinking
    (not shown) of skyrmion when it is very close to the boundary.
    The solid brown curve is Eq.~\eqref{potential_well} with
    $R=20~\hbox{nm}$, $R'=4.5~\hbox{nm}$, and
    $U_{1}=0.15~\hbox{eV}$ obtained from the micromagnetic
    simulations. Inset: three representative
    skyrmion structures in different regions.
    }\label{fig2}
\end{figure}

To have a better understanding of the physics associated
with the skyrmion polarity reversal, we numerically compute
the potential energy $U(\mathbf{X})$ of the skyrmion,
defined as the difference between the energy of a
skyrmion/twisted skyrmion
at $\mathbf{X}$ and the energy $U_0$ of a skyrmion at
the center of the left domain ($X=-150$ nm). Two
representative numerical results of $U(X)$ for
$J_{\text{AFM}}/J_{\text{FM}}=1$ and $3$ are shown by
circles and squares, respectively, in Fig.~\ref{fig2}.
We find that the behavior of the potential can be divided
into three regimes I, II, and III [shown in the inset
of Fig.~\ref{fig2}]. Regime I corresponds to the case
that the skyrmion is in the left domain, in which $U$
increases when $|X|$ decreases. The boundary exerts a 
repulsive force which balances the
Magnus force, and thus suppresses the skyrmion Hall effect. 
Therefore, if we put more skyrmions in the double track, 
all those skyrmions will move synchronously parallel with 
the boundary \cite{supp}. The potential $U$ keeps
increasing until it goes very close to the boundary, 
i.e., $-20~\hbox{nm}<X\leq 0~\hbox{nm}$.
Driven by the large electric current, the skyrmion is
able to overcome the energy barrier and to step
into the regime II ($|X|\leq 20~\hbox{nm}$), forming
a twisted skyrmion structure, in which an attractive 
potential well with its energy minimum at $X=0$ is 
clearly observed. Because the applied current is large enough,
the twisted skyrmion is pushed out of the potential well, 
and transforms back to a normal skyrmion state with 
opposite polarity in regime III. In the following, we 
quantitatively explain the potential curves obtained in all regimes.

\emph{Method of image skyrmions.$-$}We first focus on
regime I where the normal skyrmion is approaching the
domain boundary. The situation in regime III is
identical because of symmetry. Up to now, it is still
an open question on the form of the potential/force
from the boundary in the skyrmionic community.
A number of numerical studies have been reported in
the literature \cite{Iwasaki,ZhangXC1,ZhangXC2,Yoo,Carles}
employing empirical formulae, such as the quadratic
potential \cite{Iwasaki} and the exponential one
\cite{ZhangXC1,ZhangXC2,Yoo,Carles}, without analytical
justifications.
It is thus of great importance and high interest to
know the exact form of the potential $U$ between the
skyrmion and the boundary. Here, we derive the
analytical expression of $U$, to the best of our
knowledge for the first time. Our analytical results
show that neither the harmonic nor the exponential
potential is a good description. To quantitatively
formulate the problem, we write
the magnetization profile of a skyrmion as $\mathbf{m}(\mathbf{r})
=(\cos\Phi(\phi)\sin\Theta(r),\sin\Phi(\phi)\sin\Theta(r),
\cos\Theta(r))$, where $\Phi$ depends only on angular coordinate
$\phi$ and $\Theta$ relies only on radial coordinate $r$, measured
from its center $\mathbf{X}$.

\begin{figure}[ht!]
    \centering
    \includegraphics[width=0.48\textwidth]{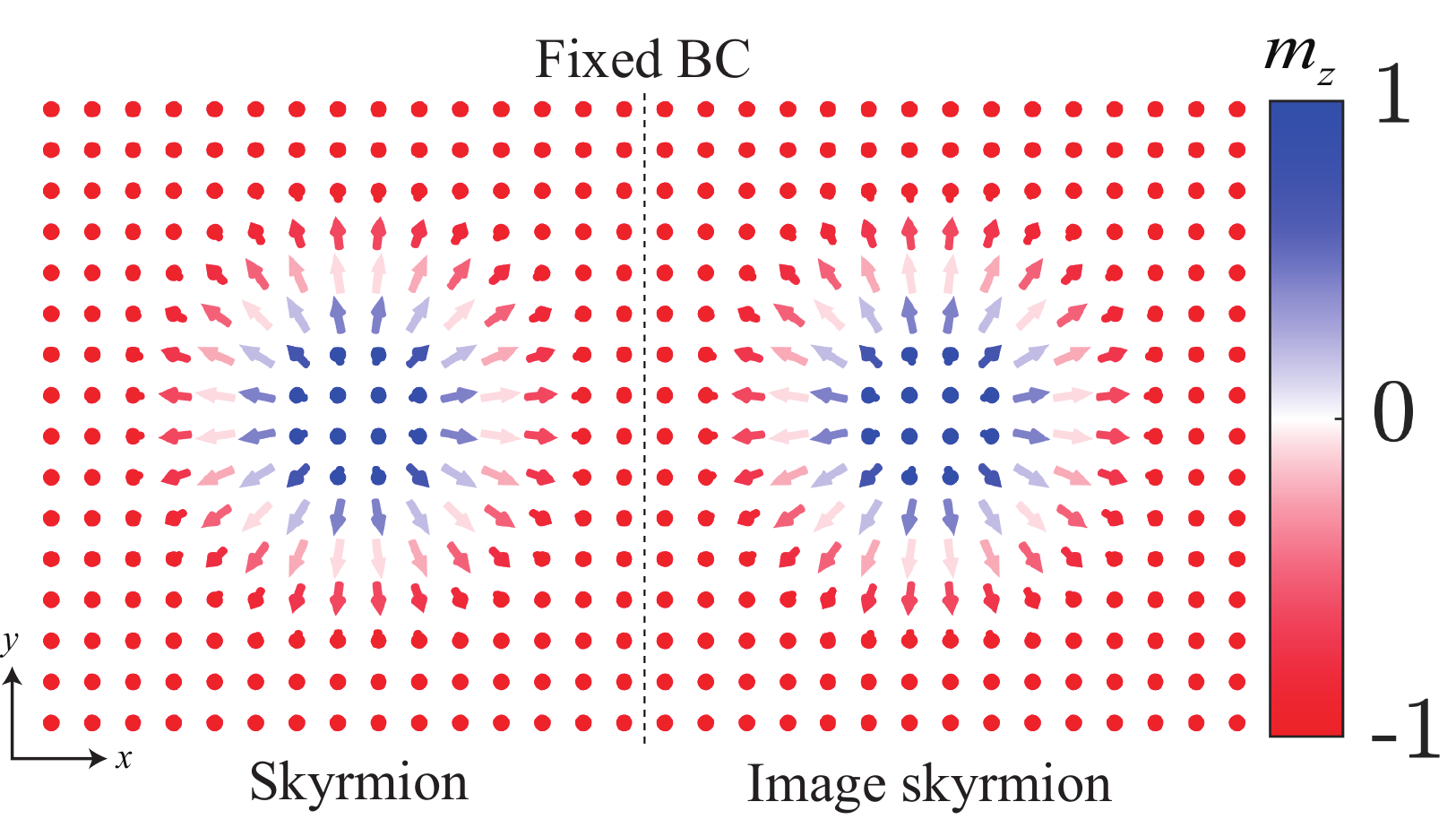}
    \caption{Schematic of the method of image in the skyrmion-boundary interaction problem. The original skyrmion is in left domain with the center coordinate $(X,Y)$ and its image locates in the right domain with the center coordinate $(-X,Y)$. Dash line represents the boundary. 
    }
    \label{fig3}
\end{figure}

The general solution for
the in-plane angle of a static symmetric skyrmion reads
$\Phi(\phi)=m\phi+\eta$, where $m$ is the vorticity
and $\eta$ is the phase angle \cite{Nagaosa}.
Remarkably, we notice that $\Phi(\phi)$ satisfies the
two-dimensional Laplace equation
\begin{equation}
\begin{gathered}
\nabla^2\Phi=0.
\end{gathered}\label{Laplace}
\end{equation}
This fact motivates us to consider the method of image to solve the skyrmion-boundary interaction problem, as long as the skyrmion has a good point-like particle nature. This is justified if the skyrmion does not significantly overlap with its image. Of course, the properties of the image skyrmion exclusively depend on the boundary condition. In our model Hamlitonian, we have introduced an interfacial AFM exchange coupling, so the edge spins in the left domain are strongly pinned by the right magnetic domain. This corresponds to the fixed boundary condition (BC) or the Dirichlet BC:
\begin{equation}
\begin{gathered}
\mathbf{m}(\mathbf{r})\cdot\mathbf{n}|_{\mathbf{r}\in\mathrm{B}}=0,
\end{gathered}\label{boundary}
\end{equation}
where $\mathbf{n}$ is the normal direction of the boundary.
We infer that the image skyrmion locates at $(-X,Y)$, assuming
the original skyrmion at $(X,Y)$, with magnetization configuration exactly being the mirror image of the
original skyrmion with respect to the boundary, to satisfy the Dirichlet BC, as shown in Fig.~\ref{fig3} (see the proof in Supplemental Materials \cite{supp}). Then the potential due to the boundary is equivalent to the interacting energy of the skyrmion$|$image-skyrmion pair. Its analytical form is obtained \cite{supp}, with very simple forms in the
following two limits
\begin{subequations}
\begin{empheq}[left={U(X)=\empheqlbrace\,}]{alignat=2}
      & \kappa_1/X^2 &\quad&\text{for }|X|\gg R,\label{sky_ima_1} \\
      & U_0-\kappa_2|X| & \quad & \text{for }|X|\gtrsim R.\label{sky_ima_2}
\end{empheq}
\end{subequations}
Here $\kappa_1=3\pi J_{\text{FM}} R^2/8$ with $R$ the radius of the skyrmion, and
$\kappa_2$ and $U_0$ are functions of $J_{\text{FM}}$
and $R$.
The agreement of our analytical formula \eqref{sky_ima_1} and \eqref{sky_ima_2} with micromagnetic simulations appears to be very well (shown in Fig.~\ref{fig2}). 
We would like to
remark that Eq.~\eqref{Laplace} is generally true for symmetric skyrmion structures.
Therefore, the method of image skyrmions can be utilized to solve a large class of skyrmion-boundary interaction problems, e.g. skyrmions in confined geometry with curved boundaries. The Dirichlet BC can also be replaced by other types of BCs, such as the Neumann BC (free BC) and the Robin BC, depending on the physical situations. Bloch skyrmion and antiskyrmion can be similarly treated as well. A thorough investigation on the application of the method of image skyrmions will be published elsewhere \cite{Prepare}.

\emph{Harmonic potential at the boundary.$-$}In regime II, the method of image skyrmion
is not valid any more because the skyrmion is not a good point-like particle and has a significant overlap with its image. However, we find that the derivation of the potential is unexpectedly simple for the twisted skyrmion at boundaries. It turns out to be a harmonic potential well \cite{supp},
\begin{equation}
\begin{gathered}
U(X)=U_1+\dfrac{k_0}{2} X^2,\text{for }|X|<R,
\end{gathered}\label{potential_well}
\end{equation}
where $U_1=U(X=0)$ is the energy minimum
of the twisted skyrmion, and $k_0=2J_{\text{FM}}R'/(a R^2)$
is an effective spring constant with $R$ ($R'$) the length of the semi-major (semi-minor)
axis of the deformed twisted skyrmion with an elliptical shape.
We find that the numerically calculated $U(X)$ ($|X|<R$) can be
well described by Eq.~\eqref{potential_well} (see Fig.~\ref{fig2}).

\begin{figure}[ht!]
    \centering
    \includegraphics[width=0.45\textwidth]{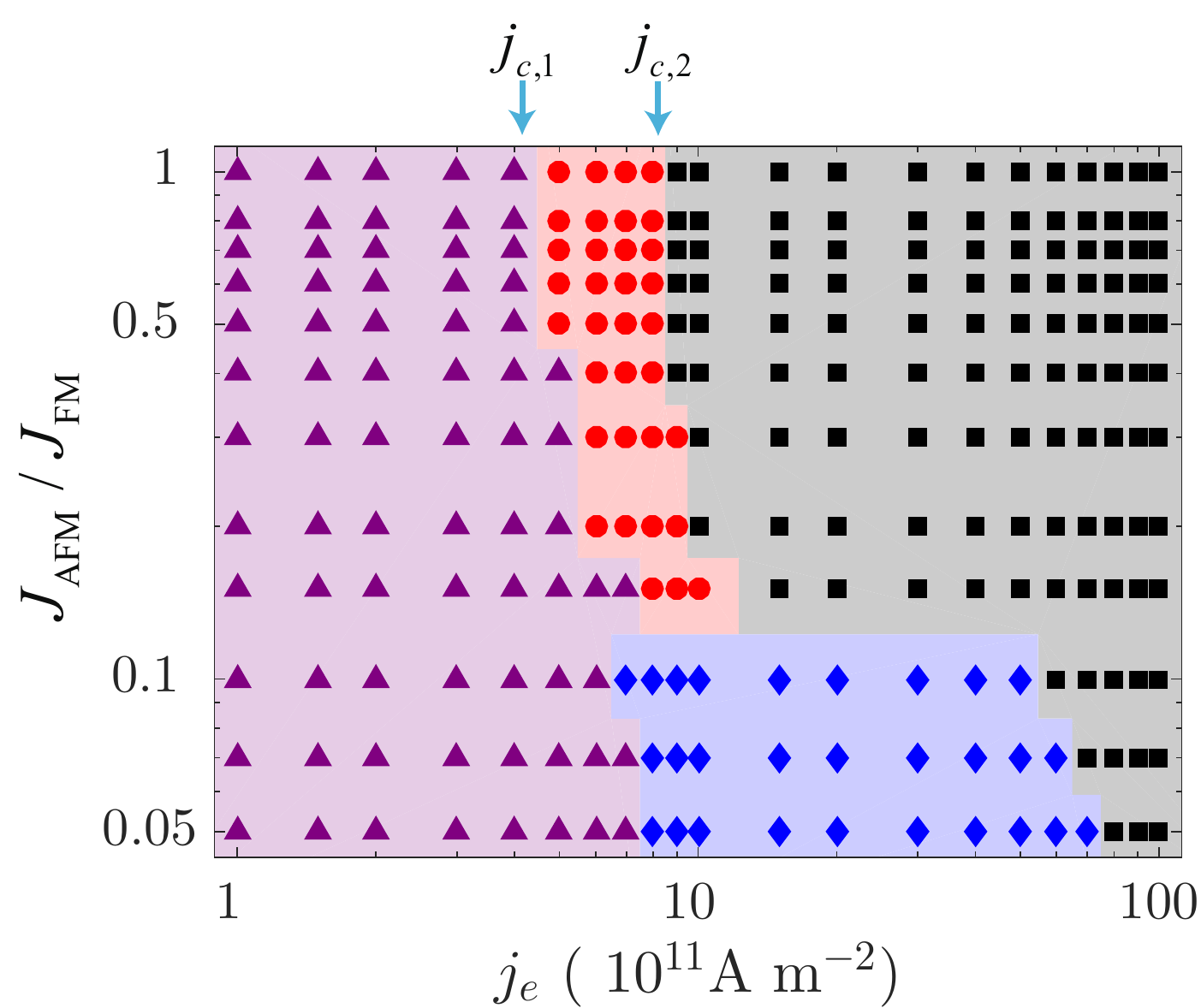}
    \caption{STAR phase diagram plotted in the
    $J_{\text{AFM}}$-$j_{e}$ plane, with black squares for ``S", red circles for ``T", blue rhombus for ``A", and purple triangles for ``R". Arrows
    label the two critical currents $j_{c,1}$ and $j_{c,2}$ analytically obtained in the large $J_{\text{AFM}}$ limit.}
    \label{fig4}
\end{figure}

\emph{Phase diagram.}$-$In previous simulations, we applied an electric current density as high as $10^{13}$ A m$^{-2}$, and identified three different skyrmion states: $p=-1$ skyrmion, twisted skyrmion, and $p=+1$ skyrmion. It is not clear what will happen if we significantly lower the current density and/or the interfacial AFM coupling. In order to obtain a complete picture, we now calculate the phase diagram by tuning the mentioned two parameters $j_{e}$ and $J_{\text{AFM}}$. Numerical results are shown in Fig.~\ref{fig4}, in which we find four
phases, labeled as ``S'' (skyrmion successfully passes through the boundary and flips its polarity),
``T'' (skyrmion is trapped at the boundary),
``A'' (skyrmion annihilates at the boundary),
and ``R'' (skyrmion is rejected by the boundary), respectively. The phase diagram is thus called STAR (see Supplemental Materials \cite{supp} for movies visualizing the skyrmion motion in different phases).
In the limit of a strong interfacial AFM coupling, the STAR phase diagram shows that both the skyrmion and the twisted skyrmion are stable. Substituting Eqs. \eqref{sky_ima_2} and \eqref{potential_well} into the Thiele's equation \cite{Thiele,Thiaville}, we are able to obtain two critical currents \cite{supp}
\begin{equation}
\begin{gathered}
j_{c,1}=\dfrac{\gamma|e|(1+\beta^2)\alpha \mathcal{D}\kappa_2}
{(16\pi^2 +\alpha\beta \mathcal{D}^2)a\mu_B}\quad\text{and}\quad
j_{c,2}=\dfrac{k_0 R}{\kappa_2}j_{c,1},
\end{gathered}\label{critical_current}
\end{equation}
for the phase transition R$\to$T and T$\to$S, respectively.

In the above equations, $\mathcal{D}=\int (\partial_x \mathbf{m})^2 d\mathbf{r}$, $\beta$ is the nonadiabatic spin torque parameter (it is set to be $0.3$ in micromagnetic simulations),
$\gamma$ is the (positive) gyromagnetic ratio, $e$ is the electron charge, and $\mu_B$ is the Bohr magneton.
Analytical formula agrees excellently with numerical results.
For a small $J_{\text{AFM}}$, the skyrmion may annihilate at
the boundary (phase ``A") when a moderate current is applied, indicating
a finite skyrmion life-time at the boundary \cite{Bessarab}. However, a large enough current can still drive the skyrmion to cross the boundary (phase ``S") before its annihilation, as long as the tunneling time is much shorter than its life-time. The existence of the phase transition A$\to$S indicates that the realization of double-track SRM actually does not rely on a very large interfacial AFM coupling.

\emph{Material considerations.}$-$The AFM coupled ferromagnetic
domains is essential to our proposal, which fortunately has been found to exist in the martensite phase
of the Heusler-type magnetic shape memory alloy \cite{Golub}.
Interestingly, a recent experiment reported the observations of
magnetic antiskyrmions in the tetragonal Heusler materials
above room temperature \cite{Nayak}. The
Heusler materials are therefore promising systems for
implementing the proposed SRM and SLGs.
In addition, the twisted skyrmions can also be
observed in the Fe-Cr-Fe multilayer and the Gd-Fe core-shell structure \cite{Peter,Camley}, the cross sections of which can support the required AFM couplings between two ferromagnetic domains.
The interfacial DMI can be generated by engineering
the surface properties of these materials and
structures \cite{Yang}.

To conclude, we reported a novel twisted skyrmion structure at the boundary separating two antiparallely aligned magnetic domains with antiferromagnetic interface coupling. We showed that skyrmions with opposite core polarities can be switched through the twisted skyrmion state.
The major obstacle of data representation to achieve the SRM and SLGs is thus removed. We
developed a method of image skyrmions to handle the boundary effect on the skyrmion, and obtained an analytical formula of boundary potential for the first time. The STAR phase diagram obtained should be useful to guide the design of the double-track SRM and SLGs. All analytical formulae are well supported by micromagnetic simulations.
Materials and structures to realize our proposals are also discussed.
We believe that our findings will shed new lights on both the fundamental sciences and the appealing applications of skyrmions.

\begin{acknowledgments}
We thank X.R. Wang for helpful discussion.
This work is supported by the National
Natural Science Foundation of China (Grants No. 11604041 and 11704060), the
National Key Research Development Program under Contract No. 2016YFA0300801,
and the National Thousand-Young-Talent Program of China.
C.W. acknowledges the financial support from the China Postdoctoral
Science Foundation (Grants No. 2017M610595 and
2017T100684) and the National Nature Science
Foundation of China under Grant No. 11704061.
X.S.W. is supported by the China Postdoctoral
Science Foundation under Grant No. 2017M612932.

H.H.Y. and C.W. contributed equally to this work.
\end{acknowledgments}

\end{document}